\documentclass[PRD,preprintnumbers,floatfix,aps,notitlepage,nofootinbib,twocolumn,showpacs,amssymb]{revtex4-1}
\usepackage{amsmath,amsfonts,bm}
\usepackage{graphicx}
\usepackage{cancel}
\usepackage[colorlinks, linkcolor={red},citecolor={blue}]{hyperref}

\begin{document}
%------------------------------------------------------------------------------------------------
%\title{Decoupling gravitational sources in general relativity: an axially %symmetric case}
%\title{Rotating black holes by gravitational decoupling}
\title{Gravitational decoupling for axially symmetric systems and rotating black holes}
\author{E. Contreras}
\email{econtreras@usfq.edu.ec}
\affiliation{Departamento de F\'isica, Colegio de Ciencias e Ingenier\'ia,
Universidad San Francisco de Quito, Quito, Ecuador.
}
\author{J. Ovalle}
\email[]{Corresponding author: jorge.ovalle@physics.slu.cz}
\affiliation{Research Centre of Theoretical Physics and Astrophysics,
	Institute of Physics, Silesian University in Opava, CZ-746 01 Opava, Czech Republic.
	%\\Departamento de F\'{\i}sica, Universidad Sim\'on Bol\'{\i}var,Apartado 89000, Caracas 1080A, Venezuela.
	%
}
\author{R. Casadio}
\email{casadio@bo.infn.it}
\affiliation{Dipartimento di Fisica e Astronomia,
Alma Mater Universit\`a di Bologna,
40126 Bologna, Italy
\\
Istituto Nazionale di Fisica Nucleare, 
Sezione di Bologna, 40127 Bologna, Italy}
\begin{abstract}
We introduce a systematic and direct procedure to generate hairy rotating black holes
by deforming a spherically symmetric seed solution.
We develop our analysis in the context of the gravitational decoupling approach,
without resorting to the Newman-Janis algorithm.
As examples of possible applications, we investigate how the Kerr black hole solution
is modified by a surrounding fluid with conserved energy-momentum tensor.
We find non-trivial extensions of the Kerr and Kerr-Newman black holes with primary hair.
We prove that a rotating and charged black hole can have the same horizon as Kerr's,
Schwarzschild's or Reissner-Nordstr\"om's, thus showing possible observational
effects of matter around black holes.
\end{abstract} 
\maketitle
%
%------------------------------------------------------------------------------------------------
%
%
\section{Introduction}
\label{intro}
Black holes (BHs) have been considered more than mere exotic solutions
of the Einstein equations for quite some time now~\cite{Gillessen:2008qv,Ghez:2008ms}.
Nonetheless, it is only very recently that their direct existence was detected,
mainly due to the spectacular results of both the LIGO~\cite{Abbott:2016blz}
and Event Horizon Telescope~\cite{Akiyama:2019cqa} collaborations.
It is also fair to mention that some ultracompact stellar models could act
as ``black-hole mimickers''~\cite{Mazur:2004fk, Cardoso:2019rvt},
although the existence of such objects would not necessarily exclude the existence
of BHs~\cite{Ovalle:2019lbs}, as it could be naively concluded.
\par
Starting with Kerr's celebrated work~\cite{Kerr:1963ud}, the interest
in BHs has increased notably, and a large number of solutions have been
found in various contexts (for some recent notable works, see
{\em e.g.}~\cite{Fan:2016hvf,Frolov:2016pav,Babichev:2017guv,
Doneva:2017bvd,Antoniou:2017acq,Herdeiro:2018wub,Silva:2017uqg} ).
Despite this diversity, in four-dimensional space-time we can group
all cases into two large groups:
i) static spherically symmetric solutions and ii) stationary rotating solutions.
(Note that, if we include non-abelian matter fields, we can also find
axisymmetric static BHs~\cite{Kleihaus:1997ic}.) 
The study of these rotating and non-rotating BH metrics, and the shadow
they produce, have been extensively explored in recent years~\cite{Bambi:2008jg,
Bambi:2010hf,Abdujabbarov:2012bn,Toshmatov:2014nya,Cunha:2015yba,
Abdujabbarov:2015pqp,Abdujabbarov:2016hnw,Younsi:2016azx,Cunha:2016wzk,
Cunha:2016bpi,Toshmatov:2017zpr,Toshmatov:2017bpx,Kumar:2018ple,
Ovgun:2018tua,Stuchlik:2018qyz,Mishra:2019trb,
Moffat:2019uxp,Contreras:2019nih,Vagnozzi:2019apd,Konoplya:2019sns,
Bambi:2019tjh,Contreras:2019cmf,Jusufi:2019caq,Allahyari:2019jqz,
Cunha:2019ikd,Konoplya:2019goy,Dymnikova:2019vuz,
Li:2020drn,Konoplya:2020hyk,Vagnozzi:2020quf,Khodadi:2020jij},
the Newman-Janis algorithm~\cite{Newman:1965tw} and its version without
complexification~\cite{Azreg-Ainou:2014pra} being tremendously useful
tools to generate rotating systems.
\par
In all cases, it is well known that the presence of matter around BHs could produce
a significant distortion of the shadow in a highly model-dependent fashion
(see {\em e.g.}~\cite{Hou:2018avu,Badia:2020pnh,Konoplya:2019sns,Konoplya:2020xam}
and references therein).
The resolution of the first BH image is not enough to support or discard
any of these models, hence it is important to study this distortion with
a minimum set of assumptions.~\footnote{Indeed, a theoretical model
with the minimum amount of assumptions is always desirable in any context}
This is precisely the topic of the present work.
Namely, we will consider a Kerr BH surrounded by an axially symmetric
``tensor-vacuum'' (analogous to the electro-vacuum and scalar-vacuum cases)
represented by a conserved energy-momentum tensor $S_{\mu\nu}$
which could account for one or more fundamental fields
(scalar, vector or tensor fields representing any phenomenologically
viable form of matter-energy, such as dark matter or dark energy).
The only restriction we require is that $S_{\mu\nu}$ satisfies either
the strong (SEC) or the dominant energy condition (DEC) in the region
outside the event horizon.
Since the Gravitational Decoupling approach (GD)~\cite{Ovalle:2017fgl,Ovalle:2019qyi} 
is precisely designed for describing deformations of known solutions of General
Relativity induced by additional sources, we will study this problem 
by first extending the GD to axially symmetric systems.  

The GD is originally based on the so-called minimal geometric deformation
(MGD)~\cite{Ovalle:2007bn,Ovalle:2020fuo} 
(for some earlier works on the MGD, see {\em e.g.}~\cite{Casadio:2012pu,Casadio:2012rf,
Ovalle:2013vna,Casadio:2013uma,Ovalle:2013xla,Ovalle:2014uwa,Casadio:2015jva,
Casadio:2015gea,Cavalcanti:2016mbe}, 
and Refs.~\cite{daRocha:2017cxu,daRocha:2017lqj,Fernandes-Silva:2017nec,Casadio:2017sze,
Fernandes-Silva:2018abr,Contreras:2018vph,Contreras:2018gzd,Contreras:2018nfg,
Panotopoulos:2018law,daRocha:2019pla,Heras:2019ibr,Rincon:2019jal,
daRocha:2020rda,Contreras:2020fcj,Arias:2020hwz,daRocha:2020jdj,Tello-Ortiz:2020euy,
daRocha:2020gee,Meert:2020sqv} for some recent applications).
The GD has been shown to be particularly useful for at least three tasks~\cite{Ovalle:2017wqi,
Gabbanelli:2018bhs,Heras:2018cpz,Estrada:2018zbh,Sharif:2018toc,Morales:2018nmq,
Sharif:2018pzr,Morales:2018urp,Estrada:2018vrl,Sharif:2018tiz,Ovalle:2018ans,
Contreras:2019fbk,Maurya:2019wsk,Contreras:2019iwm,Contreras:2019mhf,
Gabbanelli:2019txr,Estrada:2019aeh,Maurya:2019hds,Hensh:2019rtb,
Cedeno:2019qkf,Leon:2019abq,Torres:2019mee,Casadio:2019usg,
Singh:2019ktp,Maurya:2019noq,Sharif:2019mjn,Singh:2019ktp,
Abellan:2020wjw,Sharif:2020vvk,Tello-Ortiz:2020ydf,Maurya:2020rny,
Rincon:2020izv,Sharif:2020arn,Maurya:2020gjw,Zubair:2020lna}:
a) to generate new and more complex solutions from known (seed) solutions of the
Einstein field equations;
b) to systematically reduce (decouple) a complex energy-momentum
tensor $T_{\mu\nu}$ into simpler components; and
c) to find solutions in gravitational theories beyond General Relativity.
Despite the above, one of the apparent limitations of the GD is that the decoupling
of gravitational sources has only been achieved in the spherically symmetric case
so far.
One of the goals of this paper is to show that indeed the GD can be implemented
beyond spherical symmetry.
In particular, we will show how the GD can be obtained for axially symmetric systems,
which is of particular importance for the study of rotating stellar systems and BHs.
\par  
The paper is organised as follows:
in Section~\ref{Sgd}, we first review the fundamentals of the GD approach 
for a spherically symmetric system containing two sources, and then
we show in detail how to extend the GD approach for the axially symmetric case;
in Section~\ref{sec3}, we apply our results to generate the axially symmetric version
of two spherically symmetric hairy BH solutions, without implementing the
Newman--Janis algorithm.
The first solution contains a source satisfying the SEC and is an extension of the
Kerr metric, while the DEC holds for the source in the second solution, which
represents an extension of the Kerr-Newman BH;
finally, we summarize our conclusions in Section~\ref{con}.

\section{Gravitational Decoupling}
\label{Sgd}
We start this section by briefly reviewing the key aspects of the GD for spherically
symmetric systems (described in detail in Ref.~\cite{Ovalle:2019qyi}).
A particularly simple case of GD is given by the MGD~\cite{Ovalle:2007bn,Ovalle:2020fuo},  
which will guide us to introduce a GD for the axially symmetric case.
\par
We start by considering the Einstein field equations~\footnote{We use units with $c=1$
and $k^2=8\,\pi\,G_{\rm N}$, where $G_{\rm N}$ is Newton's constant.}
\begin{equation}
\label{corr2}
\tilde G_{\mu\nu}
\equiv
\tilde R_{\mu\nu}-\frac{1}{2}\,\tilde R\, \tilde g_{\mu\nu}
=
k^2\,\tilde{T}_{\mu\nu}
\ ,
\end{equation}
with a total energy-momentum tensor containing two contributions,
\begin{equation}
\label{emt}
\tilde{T}_{\mu\nu}
=
T^{\rm}_{\mu\nu}
+
S_{\mu\nu}
\ ,
\end{equation}
where $T_{\mu\nu}$ is usually associated with a known solution of General Relativity, 
whereas $S_{\mu\nu}$ may contain new fields or a new gravitational sector.
Since the Einstein tensor $\tilde G_{\mu\nu}$ satisfies the Bianchi identity, the total source
$\tilde T_{\mu\nu}$ must be covariantly conserved.
%\begin{equation}
%\nabla_\mu\,\tilde{T}^{\mu\nu}=0
%\ .
%\label{dT0}
%\end{equation} 
%
%
\subsection{Spherically symmetric case}
\label{ssc}

For spherically symmetric and static systems, the metric $\tilde g_{\mu\nu}$ can be written as
\begin{equation}
ds^{2}
=
e^{\nu (r)}\,dt^{2}-e^{\lambda (r)}\,dr^{2}
-r^{2}\,d\Omega^2
\ ,
\label{metric}
\end{equation}
where $\nu =\nu (r)$ and $\lambda =\lambda (r)$ are functions of the areal
radius $r$ only and $d\Omega^2=d\theta^{2}+\sin ^{2}\theta \,d\phi ^{2}$.
The Einstein tensor in Eq.~(\ref{corr2}) then reads
\begin{eqnarray}
\label{ec1}
\tilde G_0^{\ 0}
&=&
\frac 1{r^2}
-
e^{-\lambda }\left( \frac1{r^2}-\frac{\lambda'}r\right)
\\
\label{ec2}
\tilde G_1^{\ 1}
&=&
\frac 1{r^2}
-
e^{-\lambda }\left( \frac 1{r^2}+\frac{\nu'}r\right)
\\
\label{ec3}
\tilde G_2^{\ 2}
&=&
-\frac {e^{-\lambda }}{4}
\left(2\nu''+\nu'^2-\lambda'\nu'
+2\,\frac{\nu'-\lambda'}r\right)
\ ,
\end{eqnarray}
where $f'\equiv \partial_r f$ and $\tilde{T}_3^{{\ 3}}=\tilde{T}_2^{\ 2}$ due to the spherical symmetry.
By simple inspection, we can identify an effective density  
\begin{equation}
\tilde{\epsilon}
=
T_0^{\ 0}
+
S_0^{\ 0}
\ ,
\label{efecden}
\end{equation}
an effective radial pressure
\begin{equation}
\tilde{p}_{r}
=
-T_1^{\ 1}
-S_1^{\ 1}
\ ,
\label{efecprera}
\end{equation}
and an effective tangential pressure
\begin{equation}
\tilde{p}_{t}
=
-T_2^{\ 2}
-S_2^{\ 2}
\ .
\label{efecpretan}
\end{equation} 
Since the anisotropy  
%\begin{equation}
%\label{anisotropy}
$\Pi \equiv \tilde{p}_{t}-\tilde{p}_{r}$
%\end{equation}
usually does not vanish, the system of Eqs.~(\ref{ec1})-(\ref{ec3}) may be
viewed as an anisotropic fluid.
%~\cite{Herrera:1997plx,Mak:2001eb}.
\par
We next consider a solution to the Eqs.~\eqref{corr2} generated by the seed source $T_{\mu\nu}$
alone [that is, for $S_{\mu\nu}=0$], which we write as
\begin{equation}
ds^{2}
=
e^{\xi (r)}\,dt^{2}
-e^{\mu (r)}\,dr^{2}
-
r^{2}\,d\Omega^2
\ ,
\label{pfmetric}
\end{equation}
where 
\begin{equation}
\label{standardGR}
e^{-\mu(r)}
\equiv
1-\frac{k^2}{r}\int_0^r x^2\,T_0^{\, 0}(x)\, dx
=
1-\frac{2\,m(r)}{r}
\end{equation}
is the standard General Relativity expression containing the Misner-Sharp
mass function $m=m(r)$.
Adding the source $S_{\mu\nu}$ can then be accounted for by the deformation
of the seed metric~\eqref{pfmetric} given by
\begin{eqnarray}
\label{gd1}
\xi 
&\rightarrow &
\nu\,=\,\xi+\alpha\,g
\\
\label{gd2}
e^{-\mu} 
&\rightarrow &
e^{-\lambda}=e^{-\mu}+\alpha\,f
\ , 
\end{eqnarray}
where the parameter $\alpha$ is introduced to keep track of these deformations.
\par
By means of Eqs.~(\ref{gd1}) and (\ref{gd2}), the Einstein equations~\eqref{corr2}
split into the Einstein field equations for the seed metric~(\ref{pfmetric}), that is
\begin{equation}
\label{eins}
G_{\mu}^{\ \nu}(\xi,\mu)
=
k^2\,{T}_{\mu}^{\ \nu}
\ ,
\end{equation}
where
\begin{eqnarray}
\label{ec1pf}
&&
G_0^{\ 0}
=
\frac 1{r^2}
-
e^{-\mu }\left( \frac1{r^2}-\frac{\mu'}r\right)\ ,
\\
&&
\label{ec2pf}
G_1^{\ 1}
=
\frac 1{r^2}
-
e^{-\mu}\left( \frac 1{r^2}+\frac{\xi'}r\right)\ ,
\\
&&
\label{ec3pf}
\strut\displaystyle
G_2^{\ 2}
=
-\frac {e^{-\mu }}{4}
\left(2\xi''+\xi'^2-\mu'\xi'
+2\,\frac{\xi'-\mu'}r\right)
\ ,
\end{eqnarray}
and a second set containing the source $S_{\mu\nu}$, which reads
\begin{equation}
\label{qeins}
\alpha\,{\mathcal G}_{\mu}^{\ \nu}(\xi,\mu;f,g)
=
k^2\,{S}_{\mu}^{\ \nu}
\ ,
\end{equation}
where
\begin{eqnarray}
\label{ec1d}
{\mathcal G}_0^{\ 0}
&=&
-\frac{f}{r^2}
-\frac{f'}{r}
\ ,
\\
\label{ec2d}
{\mathcal G}_1^{\ 1}
&=&
-f\left(\frac{1}{r^2}+\frac{\nu'}{r}\right)
-Z_1
\\
\label{ec3d}
{\mathcal G}_2^{\ 2}
&=&
-\frac{f}{4}\left(2\,\nu''+\nu'^2+2\frac{\nu'}{r}\right)
-\frac{f'}{4}\left(\nu'+\frac{2}{r}\right)
-Z_2
\ ,
\quad
\end{eqnarray}
and
\begin{eqnarray}
\label{Z1}
Z_1
&=&
\frac{e^{-\mu}\,g'}{r}
\\
\label{Z2}
4\,Z_2&=&e^{-\mu}\left(2g''+g'^2+\frac{2\,g'}{r}+2\xi'\,g'-\mu'g'\right)
\ .
\end{eqnarray}
One clearly sees that the tensor $S_{\mu\nu}$ must vanish
when the metric deformations vanish ($\alpha=0$).
On assuming $g=0$, we have $Z_1=Z_2=0$ 
and Eq.~\eqref{qeins} reduces to the simpler ``quasi-Einstein'' system
of the MGD of Refs.~\cite{Ovalle:2007bn,Ovalle:2020fuo},
in which the deformation $f$ is only determined by the source $S_{\mu\nu}$ and the seed
metric~\eqref{pfmetric}.
\par
What makes the GD work is the fact that, under the transformations~\eqref{gd1} and~\eqref{gd2},
the Einstein tensor changes as
\begin{equation}
G_{\gamma} ^{\ \sigma}(\xi,\mu)
\to
G_{\gamma} ^{\ \sigma}(\nu,\lambda)
=
G_{\gamma}^{\ \sigma}(\xi,\mu)
+
\alpha\,{\mathcal G}_{\gamma} ^{\ \sigma}(\nu,\lambda)
\ .
\label{deco}
\end{equation}
That is to say, Eqs.~\eqref{gd1} and~\eqref{gd2} yield a {\it linear}
decomposition of the Einstein tensor in the parameter $\alpha$,
like the two sources add linearly in the r.h.s.~of Eq.~\eqref{corr2}.
We therefore expect that a similar GD can be introduced for any given space-time,
independently of its symmetries, if we can implement a linear decomposition
for the Einstein tensor of the form in Eq.~\eqref{deco}.
A natural application is then to consider axially symmetric systems. 	
\subsection{Axially symmetric case}
\label{asc}
Let us start with the simplest generic extension of the Kerr metric, given by the Gurses-Gursey metric~\cite{Gurses:1975vu}
\begin{eqnarray}
\label{kerrex}
ds^{2}
&=&
\left[1-\frac{2\,r\,\tilde{m}(r)}{\tilde{\rho}^2}\right]
dt^{2}
+
\frac{4\, \tilde{a}\, r\,\tilde{m}(r)\, \sin^{2}\theta}{\tilde{\rho}^{2}}
\,dt\,d\phi
\nonumber
\\
&&
-
\frac{\tilde{\rho}^{2}}{\tilde{\Delta}}\,dr^{2}
-
\tilde{\rho}^{2}\,d\theta^{2}
-
\frac{\tilde{\Sigma}\, \sin^{2}\theta}{\tilde{\rho}^{2}}\,d\phi^{2}
\ ,
\end{eqnarray}
with
\begin{eqnarray}
\tilde{\rho}^2
&=&
r^2+\tilde{a}^{2}\cos^{2}\theta
\label{f0}
\\
\tilde{\Delta}
& = &
r^2-2\,r\,\tilde{m}(r)
+\tilde{a}^{2}
\label{f2}
\\
\tilde{\Sigma}
& = &
\left(r^{2}+\tilde{a}^{2}\right)^{2}
-\tilde{a}^{2}\, \tilde{\Delta}\, \sin^{2} \theta
\label{f3}
\end{eqnarray}
and 
\begin{equation}
\label{a}
\tilde{a}\,=\,\tilde{J}/\tilde{M}\ ,
\end{equation}
where $\tilde{J}$ is the angular momentum and $\tilde{M}$ the total mass of the system.
Note that the line-element~\eqref{kerrex} reduces to the Kerr solution
when the metric function $\tilde{m}=\tilde{M}$.
Moreover, when $\tilde{a}=0$, we obtain the Schwarzschild-like metric in Eq.~\eqref{metric} with
\begin{equation}
e^{\nu}
=
e^{-\lambda}
=
1-\frac{2\,\tilde{m}(r)}{r}
\ .
\end{equation}
Hence, the correspondence between the metrics~\eqref{metric} and~\eqref{kerrex}
is clear, with the latter being a rotational version of a Kerr-Schild spherically symmetric
space-time (see
{\em e.g.}~Refs.~\cite{Guendelman:1996pg,Dymnikova:1999cz,Dymnikova:2003vt,Jacobson:2007tj}).
Although Eq.~\eqref{kerrex} is not the most general axially symmetric line element,
it can be used to describe rotating compact objects, like BHs and gravastars,
among many others.
\par
The components of the Einstein tensor for the metric~\eqref{kerrex} read
\begin{eqnarray}
\label{ec1a}
\tilde{G}_0^{\ 0}
&=&
2\,\frac{r^4+\left(\rho^2-r^2\right)^2+\tilde{a}^2\left(2\,r^2-\rho^2\right)}{\rho^6}
\,\tilde{m}'
\nonumber
\\
&&
-\frac{r\,\tilde{a}^2\,\sin^{2}\theta }{\rho^4}
\,\tilde{m}''
\\
\label{ec2a}
\tilde{G}_1^{\ 1}
&=&
2\,\frac{r^2}{\rho^4}
\,\tilde{m}'
\ ,
\\
\label{ec3a}
\tilde{G}_2^{\ 2}
&=&
2\,\frac{\rho^2-r^2}{\rho^4}
\,\tilde{m}'
+\frac{r}{\rho^2}
\,\tilde{m}''
\\
\label{ec4a}
\tilde{G}_3^{\ 3}
&=&
2\,\frac{2\,r^2\left(\rho^2-r^2\right)+\tilde{a}^2\left(\rho^2-2\,r^2\right)}{\rho^6}
\,\tilde{m}'
\nonumber
\\
&&
+\frac{r\left(\tilde{a}^2+r^2\right)}{\rho^4}
\,\tilde{m}''
\ ,
\\
\label{ec5a}
\tilde{G}^{\ 3}_{0}
&=&
2\,\frac{\tilde{a}\left(2\,r^2-\rho^2\right)}{\rho^6}
\,\tilde{m}'
-\frac{\tilde{a}\,r}{\rho^4}
\,\tilde{m}''
\ . 
\end{eqnarray}
The key observation now is that this Einstein tensor is \textit{linear}
in derivatives~\footnote{Of course, when $\tilde m=M$ is constant,
the Einstein tensor $\tilde G_{\mu\nu}=0$, since Eq.~\eqref{kerrex} is the vacuum Kerr metric.}
of the mass function $\tilde{m}(r)$, whereas the rotational parameter $\tilde{a}$
appears in a convoluted form. 
Any linear decomposition of the mass function,
\begin{equation}
\tilde{m}
=
m(r)+\alpha\,m_s(r)
\ , 
\end{equation}
will therefore generate a linear decomposition of the Einstein tensor of the form
in Eq.~\eqref{deco} with ${\mathcal G}_{\gamma}^{\ \sigma}=G_{\gamma}^{\ \sigma}$,
provided the rotational parameter $\tilde a$ is left unaffected, that is
\begin{equation}
\tilde G_{\gamma}^{\ \sigma}(\tilde m,\tilde a)
=
G_{\gamma}^{\ \sigma}(m,\tilde a)
+
\alpha\,G_{\gamma}^{\ \sigma}(m_s,\tilde a)
\ .
\label{decoA}
\end{equation}
\par
Like for the spherically symmetric case in Section~\ref{ssc},
we will assume that the mass functions $m$ and $m_s$ are generated by the
energy-momentum tensor $T_{\mu\nu}$ and $S_{\mu\nu}$ in Eq.~\eqref{emt}, respectively.
It is convenient to introduce the tetrads~\cite{Misner:1974qy}
\begin{eqnarray}
\tilde{e}^{\mu}_{t}
&=&
\frac{\left(r^{2}+\tilde{a}^{2},0,0,\tilde{a}\right)}{\sqrt{\rho^{2}\Delta}}
\ ,
\qquad
\tilde{e}^{\mu}_{r}
=
\frac{\sqrt{\Delta}\left(0,1,0,0\right)}{\sqrt{\rho^{2}}}
\label{2}
\nonumber
\\
\tilde{e}^{\mu}_{\theta}
&=&
\frac{\left(0,0,1,0\right)}{\sqrt{\rho^{2}}}
\ ,
\qquad
\tilde{e}^{\mu}_{\phi}
=
-\frac{\left(\tilde{a}\sin^{2}\theta,0,0,1\right)}{\sqrt{\rho^{2}}\sin\theta}
\ ,
\label{4}
\end{eqnarray}
so that the total source $\tilde{T}_{\mu\nu}$ generating the metric~(\ref{kerrex})
can be written as
\begin{eqnarray}\label{tmunu}
\tilde{T}^{\mu\nu}
=
\tilde{\epsilon}\, \tilde{e}^{\mu}_{t}\,\tilde{e}^{\nu}_{t}
+\tilde{p}_{r}\,\tilde{e}^{\mu}_{r}\,\tilde{e}^{\nu}_{r}
+\tilde{p}_{\theta}\,\tilde{e}^{\mu}_{\theta}\,\tilde{e}^{\nu}_{\theta}
+\tilde{p}_{\phi}\,\tilde{e}^{\mu}_{\phi}\,\tilde{e}^{\nu}_{\phi}
\ ,
\end{eqnarray}
where the energy density $\tilde{\epsilon}$ and the pressures $\tilde{p}_r$, $\tilde{p}_\theta$
and $\tilde{p}_\phi$ are given by 
\begin{eqnarray}
\label{energyax}
\tilde{\epsilon}
&=&
-\tilde{p}_{r}
=
\frac{2\,r^2}{\rho ^4}\, \tilde{m}'
\\
\label{pressuresax}
\tilde{p}_{\theta}
&=&
\tilde{p}_{\phi}
=
-\frac{r }{\rho ^2}\,\tilde{m}''
+\frac{2\left(r^2-\rho^2\right)}{\rho ^4} \,\tilde{m}'
\ ,
\end{eqnarray}
which are also, consistently, linear in (derivatives of) the mass function.
\par
We next consider a solution to the Eq.~\eqref{corr2} for the seed source $T_{\mu\nu}$
alone, which we write as
\begin{eqnarray}
\label{kerrex2}
ds^{2}
&=&
\left[1-\frac{2\,r\,{m}(r)}{\rho^2}\right]
dt^{2}
+\frac{4\, a\, r\,{m}(r) \sin^{2}\theta}{\rho^{2}}
\,dt\,d\phi
\nonumber
\\
&&
-\frac{\rho^{2}}{\Delta}\,dr^{2}
-\rho^{2}d\theta^{2}-\frac{\Sigma \sin^{2}\theta}{\rho^{2}}d\phi^{2}
\ ,
\end{eqnarray}
where the expressions for $\rho$, $\Sigma$ and $\Delta$
are the same as those in Eqs.~\eqref{f0}-\eqref{f3} but contain
$m$ and $a$ instead of $\tilde m$ and $\tilde a$.
The addition of the second source $S_{\mu\nu}$ can then be accounted
for by the GD of the seed metric~\eqref{kerrex2} given by
\begin{equation}
\label{gd3}
m(r)
\to
\tilde{m} = m(r)+\alpha\,m_s(r)
\ , 
\end{equation}
with the parameter $\alpha$ introduced to keep track of the deformation as usual.
In order to achieve the decoupling~\eqref{decoA} of Eqs.~\eqref{ec1a}-\eqref{ec5a},
we must also demand 
\begin{equation}
\label{a2}
\tilde{a}
=
a
=
a_s
\  ,
\end{equation}
that is to say, the length-scales $a$ and $a_s$ associated respectively with the sources
$T_{\mu\nu}$ and $S_{\mu\nu}$ must be and remain equal.
Finally, notice that the mass deformation~\eqref{gd3} corresponds to the particular 
metric deformation $f$ in Eq.~\eqref{gd2} given by
\begin{equation}
\label{fms}
f(r)
=
-\frac{2\,m_s(r)}{r}
\ .
\end{equation}
\par
Unlike the general GD for the spherically symmetric case, 
Eqs.~\eqref{gd3} and~\eqref{a2} split the Einstein equations~\eqref{corr2}
in two equal sets:
A) one is given by Einstein field equations with the energy-momentum tensor $T_{\mu\nu}$,
that is
\begin{equation}
\label{ec1aa}
G_{\mu}^{\ \nu}(m,a)
=
k^2\,T_{\mu}^{\ \nu}
\ .
\end{equation}
whose solution is the seed metric~\eqref{kerrex2};
B) the second set contains the source $S_{\mu\nu}$ and reads
\begin{equation}
\label{ec1ad}
\alpha\,G_{\mu}^{\ \nu}(m_s,a)
=
k^2\,S_{\mu}^{\ \nu}
\ .
\end{equation}
whose solution has the same form as the one in Eq.~\eqref{kerrex2} but with $m(r)\to\alpha\,m_s(r)$.
In this case, we conclude that the two sources $T_{\mu\nu}$ and $S_{\mu\nu}$ can be decoupled
by means of the metric deformations~\eqref{gd3} and the length-scale invariant condition~\eqref{a2}.
Notice that the energy and pressures in Eqs.~\eqref{energyax} and~\eqref{pressuresax}
can be written as~\cite{Azreg-Ainou:2014pra}
\begin{eqnarray}
\tilde{\epsilon}
&=&
\epsilon+\alpha\,\epsilon_S
\\
\tilde{p}_i
&=&
p_i+\alpha\,p_{Si}
\qquad
\left(i=r,\theta,\phi\right)
\ ,
\end{eqnarray}
where $\epsilon_S$ and $p_{Si}$ are the energy and pressures of the source $S_{\mu\nu}$.
Finally, we see that for $a=0$ the sets~\eqref{ec1aa} and~\eqref{ec1ad}
reduce to those in~\eqref{eins} and~\eqref{qeins} respectively.
We want to emphasize once again that this procedure is exact and  does not require a perturbative
expansion in the parameter $\alpha$. 
\subsubsection{Strategy}
\label{protocol}
We can now detail our scheme to generate new axially symmetric metrics
from known solutions of the Einstein field equations:
\begin{enumerate}
\item
Consider the field equations $\tilde{G}_{\mu\nu}=k^2\,\tilde{T}_{\mu\nu}$ 
which determine the metric $\tilde{g}_{\mu\nu}(\tilde{m},\tilde{a})$ in Eq.~\eqref{kerrex},
where $\tilde{T}_{\mu\nu}=T^{1}_{\mu\nu}+\ldots+T^{n}_{\mu\nu}$,
and $T^{i}_{\mu\nu}$ is the energy-momentum tensor of the $i^{\rm th}$
gravitational source.
\item 
Solve ${G}^i_{\mu\nu}=k^2\,{T}^i_{\mu\nu}$ for each $T^{i}_{\mu\nu}$
to find their respective $g^{i}_{\mu\nu}(m_i, a_i)$ in Eq.~\eqref{kerrex2}
[of the same form as~\eqref{kerrex}],
namely
\begin{eqnarray}
{G}_{\mu\nu}^{1}
&=&
k^2\,{T}_{\mu\nu}^{1}
\quad
\Rightarrow
\quad
g^{1}_{\mu\nu}(m_1, a_1)
\nonumber 
\\
&\vdots &
\nonumber
\\
{G}_{\mu\nu}^{n}
&=&
k^2\,{T}_{\mu\nu}^{n}
\quad
\Rightarrow
\quad
g^{n}_{\mu\nu}(m_n, a_n)
\ .
\nonumber
\end{eqnarray}
\item
The solution $\tilde{g}_{\mu\nu}(\tilde{m},\tilde{a})$ in Eq.~\eqref{kerrex}
of the original problem $\tilde{G}_{\mu\nu}=k^2\,\tilde{T}_{\mu\nu}$ is obtained
by setting
\begin{eqnarray}
\tilde{m}
&=&
m_1+\ldots+ m_n
\nonumber
\\
\tilde{a}
&=&
a_1 = . . . = a_n
\nonumber
\end{eqnarray}
in the line-element~\eqref{kerrex}.
\end{enumerate}
The previous scheme can be simplified even further by noting that step 2.~actually 
amounts to computing just the mass functions $m_i=m_i(r)$.
We can thus do that for $a_i=0$ for each ${G}^i_{\mu\nu}=k^2\,{T}^i_{\mu\nu}$,
{\em i.e.}~solve the spherically symmetric cases 
\begin{equation}
{G}_{\mu\nu}^{i}=k^2\,{T}_{\mu\nu}^{i}
\quad
\Rightarrow
\quad
g^{i}_{\mu\nu}(m_i, a_i=0)
\end{equation}
and generate the axially symmetric version by plugging
the mass function $\tilde{m}=m_1 + . . . + m_n$ into Eq.~\eqref{kerrex} with
the asymptotic angular momentum parameter $\tilde a$ of choice.
\subsubsection{Decoupling Einstein-Maxwell}
With the aim of testing the consistency of our approach,
let us consider a well-known case, namely the Einstein-Maxwell system.
In particular, we will consider the axially symmetric electro-vacuum,
for which the result must be the well-known Kerr-Newman solution. 
\par
Following our strategy, we start by identifying the sources
${T}_{\mu\nu}^{1}={T}_{\mu\nu}=0$ and ${T}_{\mu\nu}^{2}={S}_{\mu\nu}$
of relevance for the case at hand, that is
\begin{equation}
\tilde{T}_{\mu\nu}
=
\cancelto{0}{T}_{\mu\nu}+S_{\mu\nu}
\ ,
\end{equation} 
where
\begin{equation}
\label{max}
S_{\mu\nu}
=
\frac{1}{4\,\pi}
\left(F_{\mu\alpha}\,F^{\alpha}_{\ \nu}
+\frac{1}{4}\,g_{\mu\nu}\,F_{\alpha\beta}\,F^{\alpha\beta}
\right)
\end{equation} 
is the Maxwell tensor. 
\par
Next we solve the Einstein equations for each source separately, 
in the particularly simple case $a_1=a_2=0$.
For the the vacuum $T_{\mu\nu}=0$, we find the Schwarzschild solution
with mass 
\begin{equation}
m_1
=
M
\ .
\end{equation}
For the source $S_{\mu\nu}$, we find the Reissner-Nordstr\"{o}m solution,
whose mass function is given by
\begin{equation}
\label{mM2}
m_s(r)
=
A-\frac{Q^2}{2\,r}
\ ,
\end{equation}
where $A$ and $Q$ are integration constants, with $Q$ eventually identified
as the electric charge. 
\par
The total mass function is given by
\begin{eqnarray}
\label{mM3}
m(r)
=
M+A-\frac{Q^2}{2\,r}
\equiv
{\cal M}-\frac{Q^2}{2\,r}
\ ,
\end{eqnarray}
which, plugged into he metric~\eqref{kerrex}, yields the well-known Kerr-Newman
solution with
\begin{equation}
\tilde \Delta
=
r^2
-2\,r\,{\cal M}
+{a}^{2}
+Q^2
\ .
\end{equation}
We see that the method is straightforward, and that we do not need to use
the Newman-Janis algorithm to map the spherically symmetric solution into
the axially symmetric one.
\section{Rotating black hole solutions}
\label{sec3}
In a recent paper~\cite{Ovalle:2020kpd}, we developed a new method
to generate spherically symmetric hairy BHs by imposing a minimal set
of requirements consisting of i) the existence of a well defined event horizon,
and ii) the SEC or DEC for the hair outside the horizon.
In particular, we considered a Schwarzschild BH surrounded by a
spherically symmetric ``tensor-vacuum'' represented by a conserved
energy-momentum tensor $S_{\mu\nu}$, which is dealt with
as explained in Section~\ref{ssc}.
We will here use those solutions as seeds to generate axially symmetric
systems according to our strategy in Section~\ref{protocol}.
\subsection{Extended Kerr solution}
\label{sec seed}
\par
When we demand that $S_{\mu\nu}$  satisfies the SEC in the region
outside the event horizon, we found the extended Schwarzschild BH
metric
\begin{eqnarray}
\label{strongBH}
e^{\nu}
=
e^{-\lambda}
&=&
1-\frac{1}{r}\left(2\,M+\ell_0\right) +\alpha\,e^{-r/M}
\nonumber
\\
&=&
1-
\frac{2\,\mathcal{M}}{r}
+\alpha\,e^{-r/(\mathcal{M}-\ell_0/2)}
\ ,
\end{eqnarray}
where $\ell_0=\alpha\,\ell$ measures the increase of entropy
from the minimum Schwarzschild value $S=4\,\pi\,{M^2}$ caused by the hair,
and must satisfy 
\begin{equation}
\label{ineql}
\ell_0
\leq
2\,{\cal M}
\equiv
\ell_{\rm K}
\end{equation}
in order to ensure asymptotic flatness. 
\par
In order to extended this metric to the axially symmetric case, we
just need to identify the mass function from Eq.~\eqref{strongBH},
that is
\begin{equation}
\label{massBH}
\tilde{m}
=
{\cal M}-\alpha\frac{r}{2}\,e^{-r/(\mathcal{M}-\ell_0/2)}
\ ,
\end{equation}
which we then plug into the metric~\eqref{kerrex}.
This yields 
\begin{equation}
\label{Delta2}
\tilde \Delta
=
r^2+a^2-2\,r\,{\cal M}+\alpha\,r^2\,e^{-r/(\mathcal{M}-\ell_0/2)}
\ .
\end{equation}
The equation determining the horizon $r=r_{\rm H}$ of the metric~\eqref{kerrex}
is given by $0=\tilde g^{rr}\sim\tilde\Delta$, which yields
\begin{equation}
\label{haxi}
r_{\rm H}^2+a^2-2\,{\cal M}\,r_{\rm H}+\alpha\,r_{\rm H}^2\,e^{-r_{\rm H}/(\mathcal{M}-\ell_0/2)}
=
0
\ .
\end{equation}
We see that the Kerr horizon 
\begin{equation}
\label{kerrh}
r_{\rm Kerr}
=
{\cal M}+\sqrt{{\cal M}^2-a^2}
\end{equation}
is recovered for $\alpha=0$ and also when the inequality~\eqref{ineql}
is saturated. 
\begin{figure*}[ht!]
\centering
\includegraphics[width=0.5\textwidth]{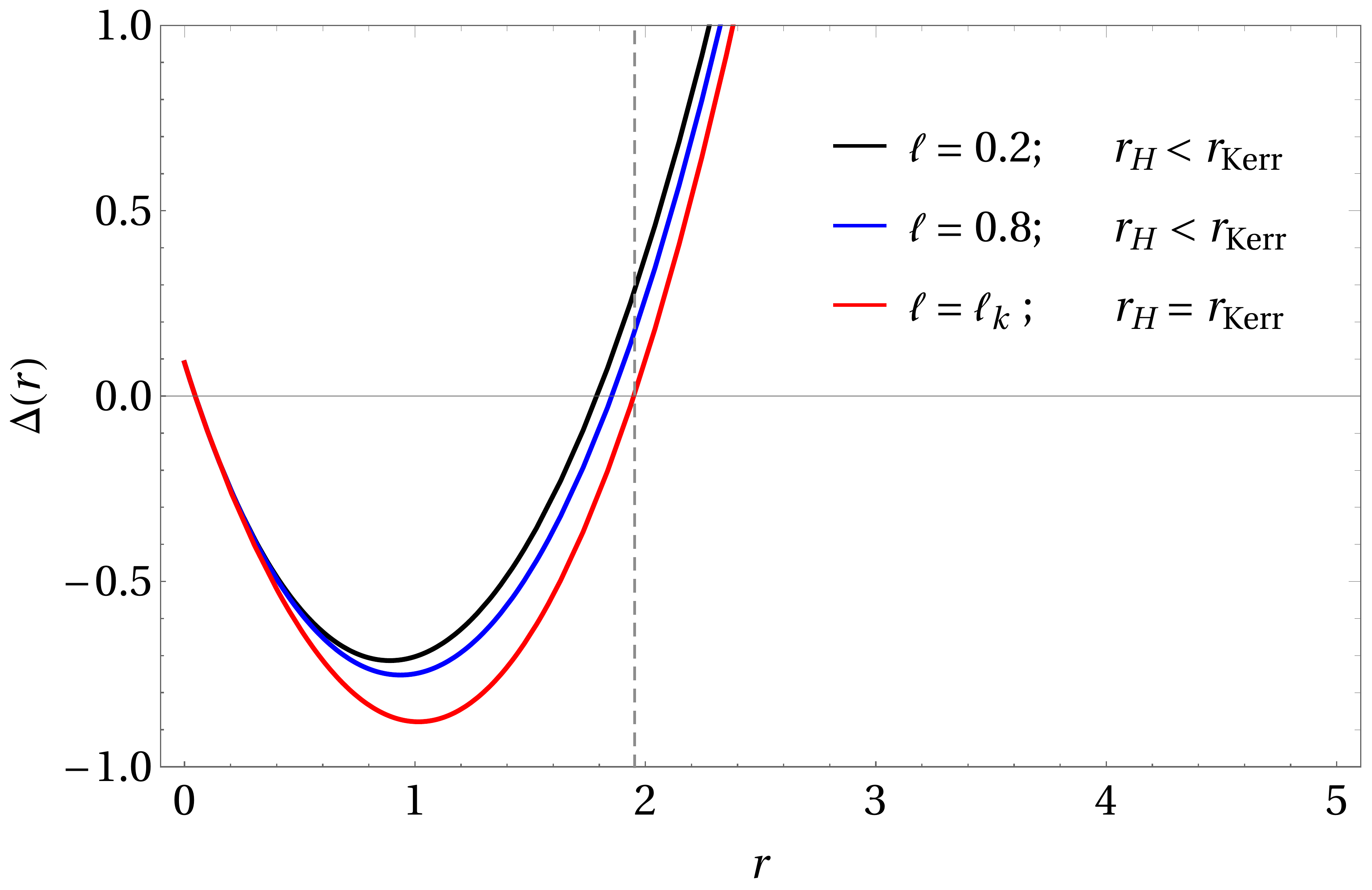}  \
\includegraphics[width=0.32\textwidth]{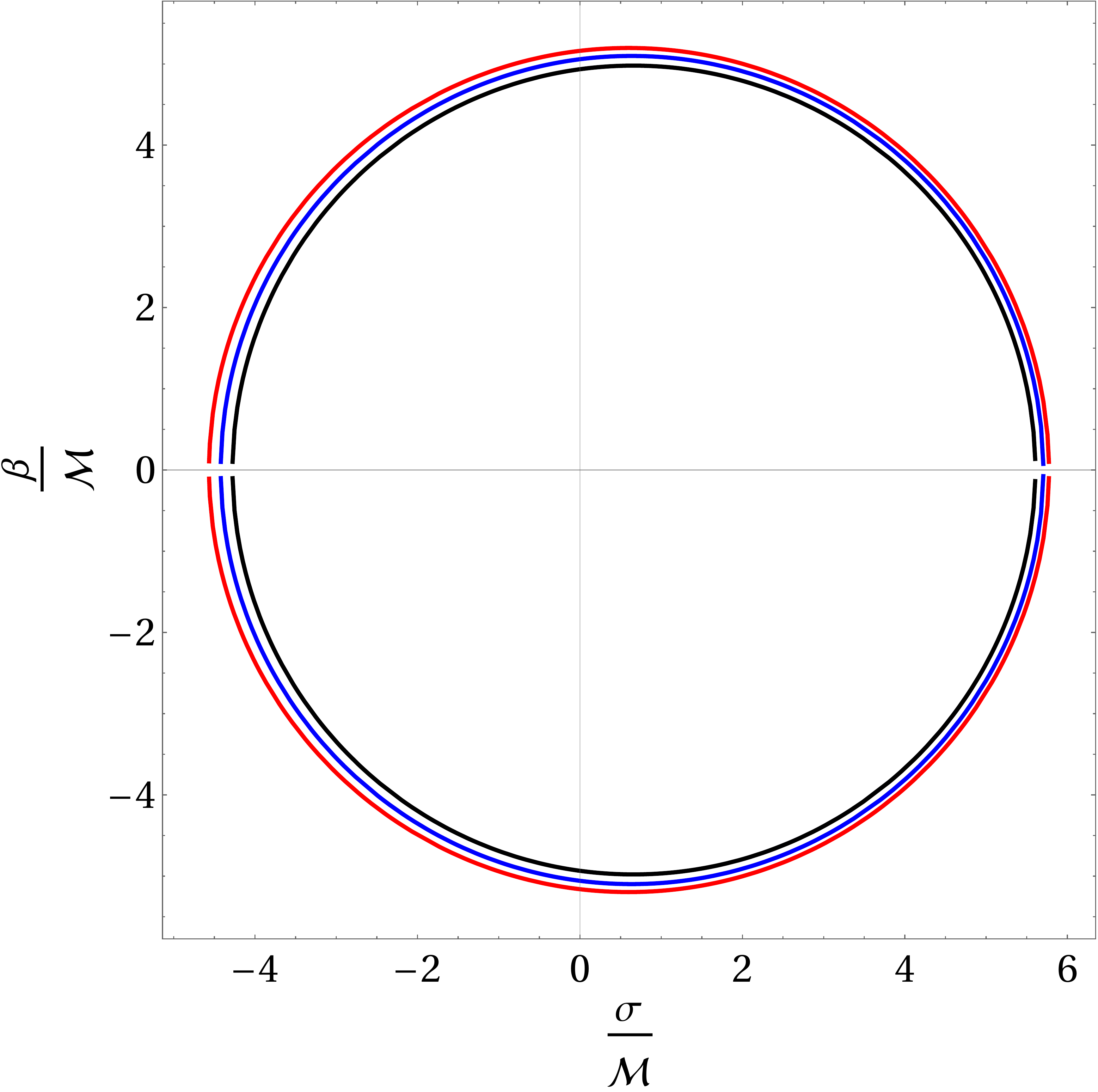}  \
\caption{Extended Kerr solution:
function $\tilde\Delta$ (left panel) and silhouette (right panel) of the shadow cast for different values
of $\ell$ with $\alpha=0.6$, $a=0.3$ and ${\cal M}=1$. 
The Kerr horizon $r_{\rm Kerr}$ here corresponds to the saturated case of Eq.~\eqref{ineql}.}
\label{fig1}
\end{figure*}
\par
The function~\eqref{Delta2} is plotted in Fig.~\ref{fig1} for a few values of $\ell_0$
at fixed $\alpha$ and $a$.
We see that the horizon shifts to larger radii when $\ell_0$
increases, reaching a maximum value corresponding to the Kerr horizon
for $\ell_0=\ell_{\rm K}$.
The silhouette of the BH is also shown in Fig.~\ref{fig1} where $\sigma$ and $\beta$
are the usual celestial coordinates (see~Appendix~\ref{apen}).
\par
We conclude that the metric~\eqref{kerrex} with the mass function~\eqref{massBH}
represents a family of rotating hairy BHs described by the parameters
$\{{\cal M},\,a,\,\ell_0\}$, where $\ell_0=\alpha\,\ell$ represents a charge associated
with primary hair.
\par
Finally, the spherically symmetric metric with components~\eqref{strongBH}
satisfies the SEC, that is
\begin{eqnarray}
\nonumber
&&
{\epsilon}_S+{p}_{Sr}+2\,{p}_{S\theta}
\geq
0
\\
\label{strong01}
&&
{\rho_S}+{p}_{Sr}
\geq
0
\\
&&
{\rho_S}+{p}_{S\theta}
\geq
0
\ .
\nonumber
%\label{strong}
\end{eqnarray}
It can be checked straightforwardly that this property is inherited by the rotating solution.
\subsection{Extended Kerr-Newman solution}
\label{dec seed}
The second case we will consider is the rotating version of the spherically symmetric
solution~\cite{Ovalle:2020kpd}
\begin{eqnarray}
\label{dominantBH}
e^{\nu}
=
e^{-\lambda}
&=&
1-\frac{2\,M+\alpha\,\ell}{r}
+\frac{Q^2}{r^2}
-\frac{\alpha\,M\,e^{-r/M}}{r}
\nonumber
\\
&=&
1-\frac{2\,\mathcal{M}}{r}
+\frac{Q^2}{r^2}
\nonumber
\\
&&
-\frac{\alpha}{r}
\left(\mathcal{M}-\frac{\alpha\,\ell}{2}\right)
e^{-r/(\mathcal{M}-\alpha\,\ell/2)}
\ ,
\end{eqnarray}
which extends a Reissner-Nordstr\"{o}m-like metric~\footnote{We remark that $Q$
is not necessarily the electric charge, but could be a tidal charge of extra-dimensional
origin or any other charge for the tensor $S_{\mu\nu}$.}
to include a conserved source $S_{\mu\nu}$ satisfying the DEC.
From~\eqref{dominantBH} we read out the mass function
\begin{equation}
\label{massBH2}
\tilde{m}
=
{\cal M}
-\frac{Q^2}{2\,r}
+\frac{\alpha}{2}
\left(\mathcal{M}-\frac{\ell_0}{2}\right)
e^{-r/(\mathcal{M}-\ell_0/2)}
\ ,
\end{equation}
which immediately yields the rotating version~\eqref{kerrex} with
\begin{equation}
\label{Delta3}
\tilde\Delta
=
r^2
+a^2
+Q^2
-2\,r\,{\cal M}
-{\alpha}\,r
\left(\mathcal{M}-\frac{\ell_0}{2}\right)
e^{-r/(\mathcal{M}-\ell_0/2)}
\ ,
\end{equation}
where again the inequality~\eqref{ineql} must hold to ensure asymptotic flatness
\par
The horizon is again determined by $\tilde\Delta=0$ or
\begin{eqnarray}
\label{haxi2}
&&
r_{\rm H}^2
+a^2
+Q^2
-2\,{\cal M}\,r_{\rm H}
\nonumber
\\
&&
=
\alpha\,r_{\rm H}
\left(\mathcal{M}-\frac{\ell_0}{2}\right)
e^{-r_{\rm H}/(\mathcal{M}-\ell_0/2)}
\ .
\end{eqnarray}
The Kerr-Newman horizon
\begin{equation}
\label{knh}
r_{\rm KN}
=
{\cal M}+\sqrt{{\cal M}^2-a^2-Q^2}
\end{equation}
is found for $\alpha=0$ (hence $\ell_0=0$)
and also when the inequality in~\eqref{ineql} is saturated. 
\begin{figure}[h!]
\centering
\includegraphics[scale=0.25]{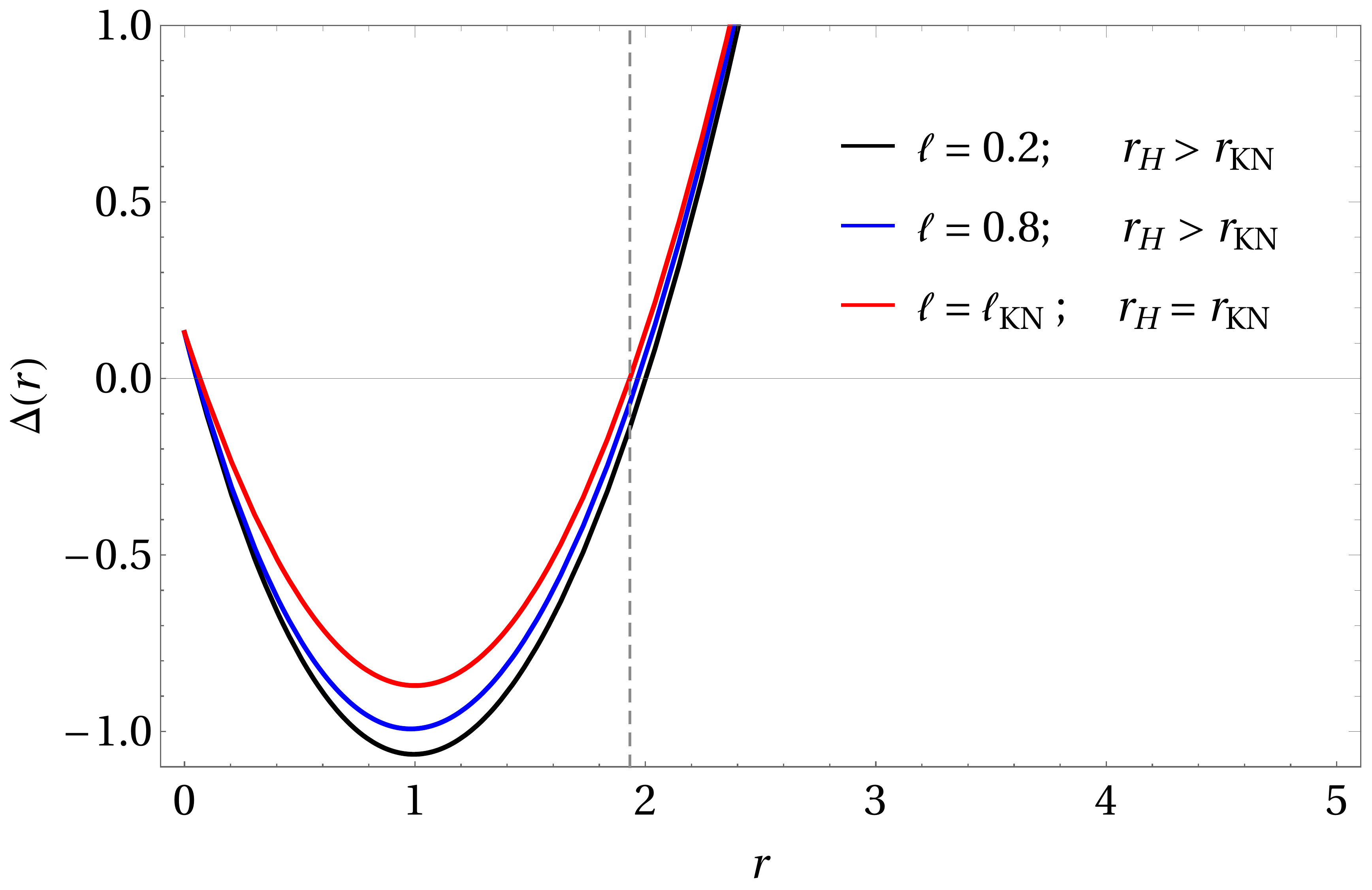}
\caption{\label{roots}
Extended Kerr-Newman solution:
function $\tilde\Delta$ for different values of $\ell$ with $\alpha=0.6$, $a=0.3$, $Q=0.2$
and ${\cal M}=1$. 
The Kerr-Newman horizon $r_{\rm KN}$ here corresponds to the saturated case of Eq.~\eqref{ineql}.}
\label{fig3}
\end{figure}
\begin{figure}[h!]
\centering
\includegraphics[scale=0.22]{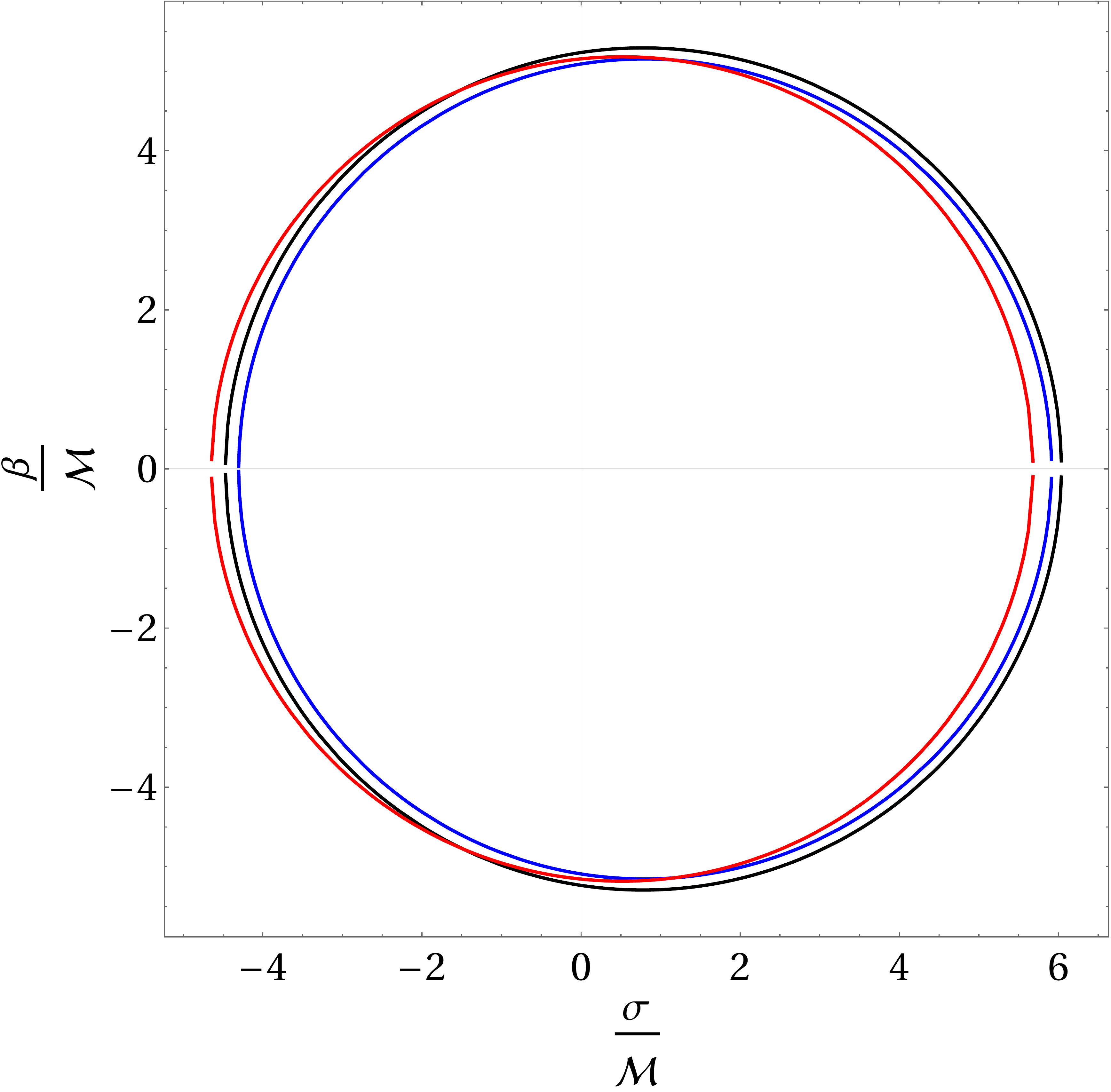}
\caption{\label{three}
Special extended Kerr-Newman solutions: 
BH shadow for (a) Schwarzschild horizon (black),
(b) Kerr horizon (blue) and (c) Reissner-Nordstr\"om
horizon (red), for $\alpha=0.9$, $\ell=0.7$, $a=0.4$, $Q=0.2$ and ${\cal M}=1$.}
\end{figure}
\par
The metric function~\eqref{Delta3} is plotted in Fig.~\ref{fig3} for given values
of $\alpha$, $a$ and $Q$.
For the range of $\ell$ shown there, the horizon shrinks to smaller radii when $\ell$
increases, reaching a minimum value corresponding to the Kerr-Newman horizon
when Eq.~\eqref{ineql} is saturated. We conclude that the metric~\eqref{kerrex} with the mass function~\eqref{massBH2}
represents rotating hairy BHs depending on the parameters $\{{\cal M},\,a,\,Q,\,\alpha,\ell_0\}$,
where $\ell_0=\alpha\,\ell$ represents a charge associated with primary hair.
\par
Like with the SEC in Section~\ref{sec seed}, one can check straightforwardly that 
the rotating metrics inherit the DEC satisfied by the spherically symmetric metric functions
in Eq.~\eqref{dominantBH}, 
\begin{eqnarray}
{\rho}_S
&\geq&
0
\label{dom1}
\\
\rho_S
&\geq&
|\tilde{p}_{Si}|
\quad
\left(i=r,\theta,\phi\right)
\ .
\end{eqnarray}
\par
It is impossible to find analytical solutions to Eq.~\eqref{haxi2}, except for some
particular cases.
Three of them are shown below.
%
%
%
%%%%%%%%%%%%%%%%%%%%%%%%%%%%%%%%%%%%%
\subsubsection*{Case~1}
If the charge and angular momentum parameters satisfy the condition 
\begin{equation}
\label{aq}
a^2+Q^2
=
\alpha\,r_{\rm H}
\left(\mathcal{M}-\frac{\ell_0}{2}\right)
e^{-r_{\rm H}/(\mathcal{M}-\ell_0/2)}
\ ,
\end{equation}
we have
\begin{eqnarray}
\label{DeltaSchw}
\tilde\Delta
&=&
r^2-2\,r\,{\cal M}
+{\alpha}\left(\mathcal{M}-\frac{\ell_0}{2}\right)
\nonumber
\\
&&
\times
\left[r_{\rm H}\,e^{-r_{\rm H}/(\mathcal{M}-\ell_0/2)}-r\,e^{-r/(\mathcal{M}-\ell_0/2)}\right]
\ ,
\end{eqnarray}
and the event horizon is located at the Schwarzschild radius $r_{\rm H}=2\,{\cal M}$.
This indicates that the source $S_{\mu\nu}$ filling the ``electro-vacuum'' produces
a screening effect on the charges $a$ and $Q$ in such a way that an external observer
will see a rotating space-time with the horizon apparently generated by a non-rotating and
neutral distribution.
%
%
%
%%%%%%%%%%%%%%%%%%%%%%%%%%%%%%%%%%%%%
\subsubsection*{Case~2}
Next, if the charge satisfies
\begin{equation}
\label{aq2}
Q^2
=
\alpha\,r_{\rm H}
\left(\mathcal{M}-\frac{\ell_0}{2}\right)
e^{-r_{\rm H}/(\mathcal{M}-\ell_0/2)}
\ ,
\end{equation}
the metric function
\begin{eqnarray}
\label{DeltaK}
\tilde\Delta
&=&
r^2+a^2
-2\,r\,{\cal M}
+{\alpha}\left(\mathcal{M}-\frac{\ell_0}{2}\right)
\nonumber
\\
&&
\times
\left[r_{\rm H}\,e^{-r_{\rm H}/(\mathcal{M}-\ell_0/2)}-r\,e^{-r/(\mathcal{M}-\ell_0/2)}\right]
\ ,
\end{eqnarray}
and the event horizon is located at $r_{\rm H}=r_{\rm Kerr}$ given in Eq.~\eqref{kerrh},
provided $a^2\le{\cal M}^2$.
This indicates a screening effect of the charge $Q$ only,
so that an external observer will detect an horizon corresponding to a
neutral distribution.
%
%
%
%
%%%%%%%%%%%%%%%%%%%%%%%%%%%%%%%%%%%%%%
\subsubsection*{Case~3}
Lastly, if the angular momentum satisfies
\begin{equation}
\label{aq3}
a^2
=
\alpha\,r_{\rm H}
\left(\mathcal{M}-\frac{\ell_0}{2}\right)
e^{-r_{\rm H}/(\mathcal{M}-\ell_0/2)}
\ ,
\end{equation}
which leads to
\begin{eqnarray}
\label{DeltaQ}
\Delta
&=&
r^2+Q^2
-2\,r\,{\cal M}
+{\alpha}\left(\mathcal{M}-\frac{\ell_0}{2}\right)
\nonumber
\\
&&
\times
\left[r_{\rm H}\,e^{-r_{\rm H}/(\mathcal{M}-\ell_0/2)}-r\,e^{-r/(\mathcal{M}-\ell_0/2)}\right]
\ ,
\end{eqnarray}
the event horizon is given by Eq.~\eqref{knh} with $a=0$, namely the Reissner-Nordstr\"om
horizon 
\begin{equation}
\label{knh}
r_{\rm RN}
=
{\cal M}+\sqrt{{\cal M}^2-Q^2}
\ ,
\end{equation}
provided of course the charge $Q^2\le {\cal M}^2$.
In this case the screening effect occurs on the rotational charge $a$.
An external observer will see a rotating BH with effective horizon corresponding
to a non-rotating charged distribution.
\par
The three screening cases above can be described collectively by the metric 
function
\begin{eqnarray}
\label{DeltaThree}
\tilde\Delta
&=&
r^2
+Z^2_i
-2\,r\,{\cal M}
+{\alpha}{L}
\left(r_{\rm H}\,e^{-r_{\rm H}/{L}}-r\,e^{-r/{ L}}\right)
\ ,
\quad
\end{eqnarray}
where $L={\cal M}-\ell_0/2$, and
\begin{equation}
\label{Z}
Z^2_i
=
\{Z^2_S,Z^2_K,Z^2_{RN}\}
=
\{0,a^2,Q^2\}
\ ,
\end{equation}
for the three effective horizons, namely, Schwarzschild, Kerr
and Reissner-Nordstr\"om, respectively.
Notice that the hair charge $\ell_0\equiv\alpha\,\ell$ in the
expressions~\eqref{aq},~\eqref{aq2} and~\eqref{aq3} is related with the
charges ${\cal M}$, $a$ and $Q$ by the Lambert ${\cal W}$ function as
\begin{equation}
\ell_{0i}
=
2\,{\cal M}
-\frac{2\,r_{{\rm H}i}}{{\cal W}\!\left(\frac{\alpha\,r_{{\rm H}i}^2}{a^2+Q^2-Z^2_i}\right)}
\ ,
\end{equation}
where the index $i$ runs on the three cases in Eq.~\eqref{Z}.
As an example, Fig.~\ref{three} shows the shadow cast in the three cases for
a given choice of parameters.
\section{Conclusions}
\label{con}
Using the GD approach and the simplest extension of the Kerr metric~\eqref{kerrex},
which could be generated by the Newman--Janis algorithm without
complexification~\cite{Azreg-Ainou:2014pra},
we have proven that the decoupling of gravitational sources in General Relativity
is possible in the axially symmetric case, as long as the metric takes the
form~\eqref{kerrex} and the asymptotic angular momentum
parameter $a$ in Eq.~\eqref{a} satisfies the critical condition~\eqref{a2}.
As a direct consequence, we provided a simple and systematic strategy to generate
axially-symmetric BHs departing from a spherically symmetric seed solution,
without implementing any variant of the Newman-Janis algorithm~\cite{Burinskii:2001bq,Dymnikova:2006wn,Smailagic:2010nv,Bambi:2013ufa}.~\footnote{Indeed, we can generate a new axially-symmetric solution by the direct  superposition of two different rotating solutions, as long as the critical condition~\eqref{a2} is satisfied. }
\par
On a formal level, our results stem from observing that the
Einstein tensor~\eqref{ec1a}-\eqref{ec5a} for the metric~\eqref{kerrex}
is linear in derivatives of the mass function.
This property could be at the heart of other known methods to generate axially
symmetric solutions of the Einstein equations from spherically symmetric solutions.
Moreover, the GD could help in investigating mass functions in non-spherically symmetric
systems~\cite{Casadio:2017uom,Casadio:2018mry,Rahim:2018unt,Giusti:2019uez,
Giusti:2019lha,Faraoni:2020vtw}.~\footnote{We plan to investigate these issues further in
separate works.} 
\par
Following the aforementioned approach, we showed how the Kerr BH, given by the
metric~\eqref{kerrex} with mass function equal to the total mass ${\cal M}$, 
is modified when a fluid with conserved energy-momentum tensor $S_{\mu\nu}$
fills the axially-symmetric vacuum.
We thus find non-trivial extensions of the Kerr BH, given by the mass
function~\eqref{massBH}, and Kerr-Newman BH, with mass
function~\eqref{massBH2}.
Both solutions can support a primary hair $\ell_0\leq2{\cal M}$,
whose impact on the silhouette of the shadow is displayed in
Figs.~\ref{fig1} and~\ref{three}. 
\par 
Finally, in the case of the extended Kerr-Newman BH, whose horizon
is found by solving Eq.~\eqref{haxi2}, we identify special cases describing
a screening effect induced by the source $S_{\mu\nu}$ on the charges $a$ and $Q$,
in such a way that an external observer would see a rotating BH with effective horizon
corresponding to i) non-rotating and neutral distribution, ii) neutral distribution, iii)
non-rotating but charged distribution.
This clearly indicates that the matter around BHs may have a significant
observational impact and separating different models of BHs could
remain a very hard task. In this respect, we notice that these results 
do not contradict the conclusions of Ref.~\cite{Gurlebeck:2015xpa} about no hairs
for BHs in astrophysical environments, since the (effective) fluid modifying the Kerr
geometry overlaps the BH horizon in our case.
\par
Although it is not the main goal of the present work, we would like to conclude
by mentioning that both rotating solutions [characterized by the metric functions
in Eqs.~\eqref{Delta2} and~\eqref{Delta3}, respectively] could be investigated
further, in particular, for possible observational constraints on the primary hairs
$\ell_0$ and $Q$.
However, this is beyond the purpose of the present work.
\subsection*{Acknowledgments}
R.C.~is partially supported by the INFN grant FLAG and his work has also been carried
out in the framework of activities of the National Group of Mathematical Physics
(GNFM, INdAM) and COST action {\em Cantata\/}. 
\begin{appendix}
\section{Null geodesics around rotating BHs}
\label{apen}
We briefly review how to study null geodesics in a rotating space-time
like the one in Eq.~\eqref{kerrex}, 
%by means of the Hamilton-Jacobi formalism,
and find the celestial coordinates describing the BH shadow.
Waves propagate along characteristic curves described by the Hamilton-Jacobi
equation
\begin{equation}
\label{jac}
\frac{\partial S}{\partial\lambda}
=
\frac{1}{2}\,g^{\mu\nu}\,\partial_{\mu}S\,\partial_{\nu}S
\ ,
\end{equation}
where $\lambda$ is a parameter along the curve and $S$ the Jacobi action.
Given the symmetries of the space-time~\eqref{kerrex}, Eq.~(\ref{jac}) is separable
and one has~\cite{Carter:1968rr}
\begin{equation}
\label{separable}
S
=
-E\,t+\Phi\,\phi+S_{r}(r)+S_{\theta}(\theta)
\ ,
\end{equation}
with $E$ and $\Phi$ being the conserved energy and angular momentum, respectively.
Replacing~\eqref{separable} in Eq.~\eqref{jac}, we obtain
\begin{eqnarray}
S_{r}
& = & 
\int\frac{\sqrt{R(r)}}{\Delta}\,dr
\nonumber
\\
\\
S_{\theta}
& = &
\int\sqrt{\Theta(\theta)}\,d\theta
\ ,
\nonumber
\end{eqnarray}
where
\begin{eqnarray}
R
&=&
\left[
(r^{2}+a^{2})\,E
-a\,\Phi
\right]^{2}
-
\Delta
\left[
Q+
(\Phi-a\,E)^{2}
\right]
\nonumber
\\
\\
\Theta
&=&
Q
-
(\Phi^{2}\,\csc^{2}\theta-a^{2}\,E^{2})\,\cos^{2}\theta
\ ,
\nonumber
\end{eqnarray}
with $Q$ the Carter constant.
\par
The (unstable) circular photon orbits are determined by $R=R'=0$, namely
\begin{eqnarray}
&&
\left(a^2-a\, \xi +r^2\right)^2
-
\left(a^2+r^2 F\right)
\left[
(a-\xi )^2+\eta
\right] 
= 0
\nonumber
\\
\\
&&
4 \left(a^2-a \,\xi +r^2\right)
-
\left[
(a-\xi )^2+\eta \right]
\left(r\, F'+2\, F\right)
= 0
\ ,
\nonumber
\end{eqnarray}
where $\xi=\Phi/E$ and $\eta=Q/E^{2}$ are the impact parameters.
Accordingly, 
\begin{eqnarray}
\xi
&=&
a
+\frac{r^2}{a}
-\frac{4 \left(a^2+r^2\, F\right)}{a\left(r\, F'+2\, F\right)}
\ ,
\nonumber
\\
\eta
&=&
\frac{r^2 \left[r^2+2\,a \left(a-\xi \right)-\left(a-\xi \right)^2 F\right]}{a^2+r^2\, F}
\ ,
\\
F
&=&
1-\frac{2\,\tilde{m}}{r}
\ ,
\nonumber
\end{eqnarray}
where $r$ is the radius of the unstable photon orbit.
\par
The apparent shape of the shadow is finally described by
the celestial coordinates~\cite{Vazquez:2003zm}
\begin{eqnarray}
\sigma
&\equiv&
\lim\limits_{r_{0}\to \infty}
\left(-r_{0}^{2}\,\sin\theta_{0}\frac{d\phi}{dr}\bigg|_{(r_{0,\theta_{0}})}\right)
\nonumber
\\
&=&
-\frac{\xi }{\sin\theta_{0}}
\label{sigmaf}
\end{eqnarray}
and
\begin{eqnarray}
\beta
&\equiv&
\lim\limits_{r_{0}\to\infty}
\left(r_{0}^{2}\,\frac{d\theta}{dr}\bigg|_{(r_{0},\theta_{0})}\right)
\nonumber
\\
&=&
\sqrt{\eta -\xi ^2 \,\cot ^2\theta_{0}+a^2\, \cos ^2\theta_{0}}
\ ,
\label{betaf}
\end{eqnarray}
where $(r_{0},\theta_{0})$ are the coordinates of the observer.
\end{appendix} 
%
%
%
%\section*{References}
\bibliography{references.bib}
\bibliographystyle{apsrev4-1.bst}
%
%
%%%%%%%%%%%%%%%%%%%%%%%%%%%%%%%%%%%%%%%%%%%%%%%%%%%%%%%%%
%\section{Acknowledgements}--------------------------------------------------------
\end{document}